# Critical Incidents for Technology Enhanced Learning in Vocational Education and Training

Observations from the field of mechanical engineering

– working draft 2019-01-13,  Adrian Wilke,  Johannes Magenheim,  Paderborn University –

*Abstract*—In this study, observations of the Vocational Education and Training (VET) in mechanical engineering companies are carried out. A Learning Management System (LMS) had been developed for the assistance in solving typical task structures, that are used for a period of three and a half years in the apprenticeship. In this study, the Critical Incident Technique (CIT) is applied for the observations. For the subsequent analysis, a classification of incidents is performed. The most important incidents as well as conclusions for Technical Enhanced Learning (TEL) in similar domains are presented.

## I. INTRODUCTION

Work-based learning in the Vocational Education and Training (VET) System in Germany has undergone a development that can be traced back to the early industrial age [2]. Today, a regular apprenticeship takes place in companies and vocational schools within the 'dual system'. International consultations for other countries are offered to implement practice-oriented VET solutions [3]. As a result of the far-reaching tradition, working methods have become established that are highly specialized on the one hand. On the other hand, standardized operations complicate the integration of innovative working methods. In everyday VET practice, printed versions of spreadsheets and textbooks are often used and advantages of the digital counterparts are not entirely utilised [4]. This study is part of a research project in which a mobile learning system to provide assistance for VET is developed and whose requirements are characterized by the following properties:

- Learning and work tasks (LWT) in the apprenticeship and the fields of mechanical engineering and metalworking comprise theoretical competencies and practical activities in diverse environments.
- An apprenticeship usually lasts 3.5 years. The entire period has to be supported.
- The mobile learning system to be developed has to cover a cross-company use, including international companies with thousands of employees and small training centers at inter-company level.

This work is part of the completed Mobile Learning in Smart Factories (MLS) project, which was funded by the German Federal Ministry of Education and Research (funding code: 01PD14009B) and implemented in cooperation with the non-profit company Nachwuchsstiftung Maschinenbau gGmbH.

### A. Related work

As a result of continuing developments in technology, teaching in all types of fields demands to utilize arising advantages and to investigate the emerging interconnections. Concerning the core areas Technical and Vocational Education and Training (TVET/VET) and Technology Enhanced Learning (TEL), the following works and their interrelations to this study provide a theoretical foundation for the comprehension of the contents afterwards.

International approaches of guided workplace learning (including supervision/mentoring/coaching) focused on practices, guidance providers, and influencing factors, were reviewed in a meta study by Mikkonen et al. [5]. Common guidance practices are the cooperation of experts and novices, explanations based on advisers expertise, reflection in form of verbal exchanges, observations and demonstrations. It was found that the entire work community including designated trainers collaborate. Positive guidance factors are the responsibility of learners, supportive relationship with a trainer, independent work of learners and the integration of theory and practice. On the contrary, instances for hindering factors are the lack of resources, a marginal role of learners, discrepancies of learning environments, and unstructured training. The majority of these factors can be confirmed by the results of this study.

The term mobile learning has become its own research field, including contributions from the fields vocational training and industrial manufacturing. For instance, Sonderegger et al. [6] focused site-specific contexts and safety instruction scenarios supported with QR-codes and beacons. Krauss et al. [7] developed a mobile application, where learners are supported by a recommendation engine for learning content, while trainers get an overview of learners progression. In contrast to the recommendation engine of that approach, we use handpicked learning contents by domain experts, which are integrated for every work phase of all tasks in the integrated task catalog. The contents themselves are, inter alia, extracts from the Mechanical and Metal Trades Handbook (47th edition in original language) [8], which is used nearly everywhere in the given VET domain in Germany. Learning contents, including technical drawings and videos, are additionally indexed and available via the integrated search engine for autonomous learning at different levels of knowledge. Trainers are free to adapt the tasks (including hierarchy and sequences subtasks, instructions, and contents) to local conditions via an integrated

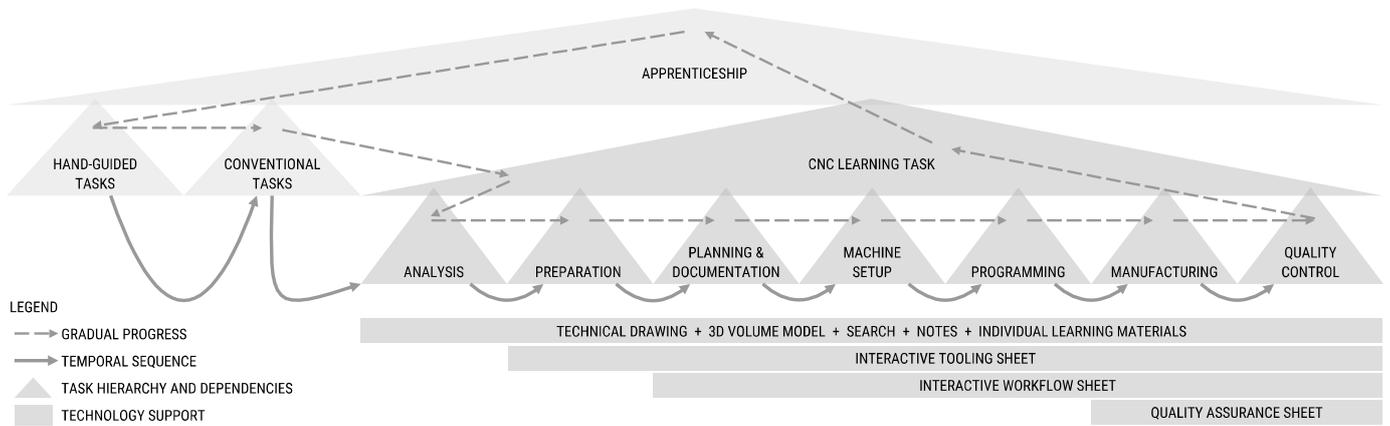

Fig. 1. Hierarchical-sequential organization of activities and technology support in individual work phases. Model based on Volpert and Hacker, see [1]

editor and to monitor the progress and results of tasks of apprentices.

Problem-solving approaches of a mechatronics expert and several lecturers using a modified epistemic plane of Legitimation Code Theory (LCT) for curriculum redesign was investigated by Wolff [9]. She visualizes the problem solving process of persons with different expertise by noting the sequence of sub-tasks on two axes, which represent the clarity of phenomenons and open or fixed approaches. In our context, LWT in general would be classified as well-known phenomenons with established (fixed) solution approaches. Although, for the evaluation of concrete solution processes of individual learners, single solution paths and the use of technology support form the core of the observations in the following with regard to develop a cross-company TEL approach. In the field of Technology Enhanced Learning, the support of online environments for Self Regulated Learning was investigated [10]. Our approach matches the following 3 most supported strategies: (1) A comparison of predefined and actually manufactured dimensions of workpieces enables a self-evaluation. (2) Setting goals for individual sub-tasks is trained through recurring instructions, which are refined for all task specifications and requirements. (3) Help seeking for contents is offered via an embedded search engine for high-quality learning contents developed by domain experts. Additionally, the concept of our learning system enforces direct face-to-face communication with persons in the community of practice inside companies [11].

For the integration of learning software into existing learning, teaching and work processes, current requirements have to be analyzed to ensure an appropriate assistance of target groups. For the realization of the software and presentation in embedded workshops, we utilized two approaches. Firstly, the Technological Pedagogical Content Knowledge (TPACK) [12] was used to focus on the overlap of the dimensions technology (here: the integration of the mobile application), pedagogy (here: focus on media didactics), and content (here: the underlying domains of VET and metalworking). Secondly, the results of a research project concerning media competence in VET [13] have been utilized to discuss most important fields for trainers and apprentices. The most important aspects from the perspective of workshop participants are autonomous and purposeful learning (.92) and to master specialized software (.83).

*B. Insight into the research project*

In the research project MLS, a mobile LMS for everyday LWT in training companies was developed. The target group of trainers is supported in the organization of training and in the allocation and evaluation of work tasks. Apprentices receive assistance in the execution of the tasks, in which the production of work pieces must be planned and executed on metal cutting machines. This takes place in the seven following work phases: (1) analysis of mechanical drawings, (2) preparation of a machine and cutting tools, (3) documentation using a work plan, (4) setup by measuring the dimensions of cutting tools and loading the machine, (5) programming the machine, (6) manufacturing, and (7) quality control. To determine requirements for the support of such tasks by means of digital media and technology, a preliminary analysis along with a formative evaluation has been carried out. We interviewed 15 apprentices [4] and 9 trainers [11] from mechanical engineering companies. The results have been integrated into the development of a mobile LMS. The LMS is based on underlying learning conditions as well as teaching-learning-objectives (predetermined in training regulations) and was developed focusing on the material, process and social aspects of instructional designs [4]. For the social aspect, classical direct face-to-face communication was preserved, as it was considered an important factor by trainers [11]. The LMS leads through the work process and provides support for the execution of the respective work phases, e.g. by 3D visualisations of technical drawings, digital work plans, and digital textbook contents adapted to current activities.

This study is based on the identified requirements and the developed LMS. For a practical evaluation of the integrated

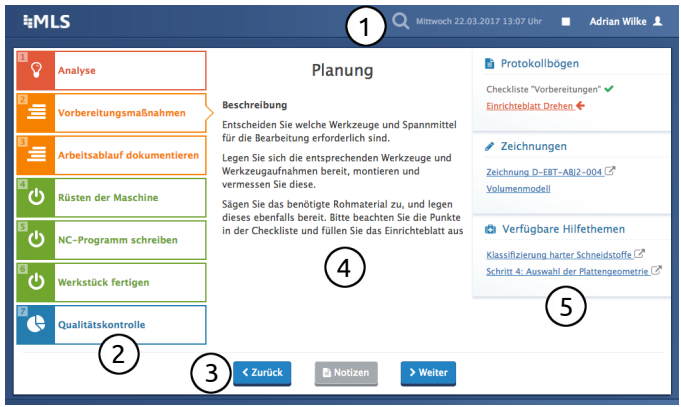

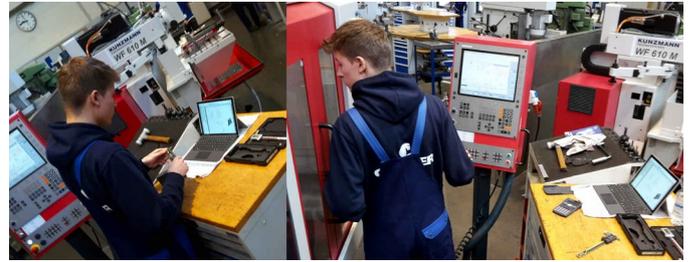

Fig. 3. Apprentice at the quality assurance of a work task

Fig. 2. Graphical user interface of a task view: (1) search and general information (2) sub-task navigation and overview (3) secondary navigation and notes (4) instructions and information (5) linked learning contents and interactive sheets

system, apprentices were observed during the execution of work tasks. The aim of this evaluation is to further improve the mobile LMS and to optimize the learning processes integrated into the workplace.

In VET and the German dual system, the model of a "complete action" is frequently referenced. It is based (and simplified) on works of the industrial psychologists Hacker and Volpert. An introduction of the model of action by Hacker and the hierarchical-sequential organization by Volpert has been translated in [1]. Fig. 1 integrates the classical perspectives of work and organizational psychology and our current approach of technology assistance. It shows a simplified model of task dependencies, sequences of activities and applied technology support for individual work phases of a typical LWT in the apprenticeship as well as the field of mechanical engineering. The dominant type of activity in VET companies and in the three-and-a-half year vocational training program are practical tasks. These typically begin with a basic training in metalworking with hand-guided tools, continue with conventional (not programmable) machine tools and lead to the production of component groups with CNC machines. Over time, tasks become more complex and fine-grained. For instance, at the beginning of VET, the setup of machines, consisting of a configuration and measuring tools to be clamped, is not necessary. The programming work phase is only required in case of the use of programmable machines and respective controls. Furthermore, the representation of tasks in Fig. 1 is simplified. In reality, recurrences between work phases arise during learning processes and situational conditions. Additionally, single tasks could be represented more specifically and would therefore be subdivided.

In the TEL approach of this paper, we provide a catalogue of tasks for changing knowledge levels of learners. The tasks are selected by trainers an assigned to apprentices. The learners access learning resources by themselves.

Tasks are sub-divided into sub-tasks that build on each other. Figure 2 shows the graphical representation of three main columns. Each sub-task consists of a navigation view to change back to previous sub-tasks, a view with instructions, and a view to display interactive elements and learning contents. The elements are designed and selected to assist in solving the objectives of the respective sub-tasks.

## II. METHODS

For the observations in this study, we applied the Critical Incident Technique (CIT) to collect decisive events during the solving of tasks. The incidents were classified and analyzed afterwards.

### A. The Critical Incident Technique (CIT)

The used CIT 'consists of a set of procedures for collecting direct observations of human behaviour in such a way as to facilitate their potential usefulness in solving practical problems' [14]. The original procedure, that we have applied, is composed of the following steps: (1) Formulate general aims of the activity to be investigated, (2) create plans and specifications, a brief statement obtained from the authorities in the field which expresses in simple terms those objectives to which most people would agree, (3) collect data in observations, (4) analyze data, and (5) interpret and report the results. In over 50 years the CIT has been applied in many disciplines, e.g. education and teaching as well as organizational learning [15].

### B. Application of CIT in the study

In this study, the CIT was executed by the following steps: (1) The general aim of work tasks has been determined as 'the autonomous production of components and sub-assemblies involving the planning, execution and checking of work'. This is based on the related official ordinance [16] and all upcoming observations are evaluated against the relevance to this general aim. (2) We prepared observational forms including the context of a task (inter alia: component specification, CNC or conventional manufacturing, work with/without the mobile LMS, learning situation). For every incident, the following data had to be recorded: (a) work phase, (b) the observed activity itself, (c) estimated positive or negative extent of the effect on the general aim, and (d) an optional explanation of the assessment. Additionally, we integrated 11 questions, that were asked after each work phase. The use of think-aloud protocols was discarded as the volume at manufacturing facilities did not allow them. (3) We collected data on 15 task executions in 10 training companies from February to October 2017. One observer was a prospective master craftsman and another observer had a background in Technology Enhanced Learning. Fig. 3 provides an impression of an observation in which the LMS and a notebook was used. (4) We analyzed the incidents and questions regarding the work phases and processes, and (5) interpreted the data. The report on the results of this assessment takes place within this paper.

TABLE I
NUMBER OF OBSERVED INCIDENTS DURING WORK PHASES

| work phase | estimated extent of the effect on the general objective (in brackets: duplicates during observations) | | | | | | | | | | | | | |
|---|---|---|---|---|---|---|---|---|---|---|---|---|---|---|
| | strongly negative | | negative | | weakly negative | | weakly positive | | positive | | strongly positive | | not specified | | total | |
| analysis | (0) | 0 | (2) | 2 | (4) | 4 | (2) | 2 | (3) | 3 | (2) | 2 | (2) | 2 | (15) | 15 |
| preparation | (1) | 1 | (11) | 9 | (17) | 14 | (7) | 6 | (4) | 4 | (0) | 0 | (5) | 5 | (45) | 39 |
| documentation | (1) | 1 | (18) | 17 | (15) | 14 | (6) | 6 | (17) | 15 | (4) | 4 | (3) | 3 | (64) | 60 |
| setup | (0) | 0 | (2) | 2 | (4) | 3 | (2) | 2 | (6) | 6 | (1) | 1 | (3) | 3 | (18) | 17 |
| programming | (0) | 0 | (3) | 3 | (6) | 6 | (2) | 1 | (15) | 13 | (0) | 0 | (5) | 5 | (31) | 28 |
| manufacturing | (1) | 1 | (19) | 17 | (18) | 16 | (20) | 20 | (25) | 22 | (6) | 6 | (7) | 7 | (96) | 89 |
| quality control | (3) | 3 | (5) | 5 | (4) | 4 | (2) | 2 | (4) | 3 | (1) | 1 | (1) | 1 | (20) | 19 |
| not specified | (0) | 0 | (0) | 0 | (0) | 0 | (0) | 0 | (1) | 1 | (0) | 0 | (2) | 2 | (3) | 3 |
| total | (6) | 6 | (60) | 55 | (68) | 61 | (41) | 39 | (75) | 67 | (14) | 14 | (28) | 28 | (292) | 270 |

*C. Classification and analysis of incidents*

Every model represents certain aspects of reality, while other aspects are relegated to be secondary or even not considered. In order to classify incidents without consequential limitations, we developed a model with which incidents can be assigned to hierarchical disjunctive top-level categories in order to subsequently investigate specific correlations across sub-categories. The model is comprised of the following top-level dimensions:

**Process:** Processes are considered as temporal sequences, that involve changes of states. For instance, established work processes typically contain a preparation, an actual implementation, and a post processing while a learning object is modified.
**Artifact:** Artifacts cover material objects as well as theoretical units. This includes, for example, technical drawings, hardware components, software elements of the LMS, and data streams.
**Social:** The social dimension concerns interactions between autonomous subjects independent of the utilized communication channel. For instance, apprentices can use communities of practice as learning resources by direct face-to-face communication.

Concerning the classification and the subsequent evaluation, on the one hand, the classification model should be kept abstract in order to draw general conclusions. On the other hand, incidents should be classified as specifically as possible. To achieve this, analysts may define sub-dimensions to create refined categories. Correlations between them can be examined, depending on the main research focus. Additionally, it is explicitly permitted to classify an incident into several categories.

The model that is used is derived from didactic design considerations of Reinmann [4], [17], which focuses on the dimensions (a) activation of learners and design of tasks, (b) mediation and design of contents, and (c) support and communication measures. Reinmann emphasizes the importance of learning objectives and also includes underlying conditions. The didactic design for the technological support in the field covered in this study is based on requirements including objectives and conditions of learning and work tasks in the apprenticeship [4]. In contrast to the dimensions of Reinmann, we focus on the classification of observed activities and their relation to (a) work- and learning processes, (b) utilized artifacts in individual situations, and (c) social interactions that take place during problem solving.

III. RESULTS

A total of 292 incidents was recorded during the 15 observations. In 7 observations, the two observers participated simultaneously. During these observations, 22 incidents were noted multiple times and have been aggregated afterwards. The aggregation was applied, if similar textual descriptions and same estimations were recorded. It resulted in a total of 270 deduplicated incidents. Table I shows a quantitative comparison of the work phases in which the incidents occurred as well as the extent of the estimated effect on the general aim (see Sec. II). The number of incidents recorded in the individual work phases was considerably influenced by the different amount of required working time. The manufacturing phase took the most time, consequently most of the incidents were collected in this phase.

TABLE II
NUMBER OF CLASSIFIED INCIDENTS IN TOTAL (IN BOLD TYPE) AND INTERSECTIONS (IN RELATIVE FREQUENCIES TO CATEGORIES ON THE LEFT)

|  | process | | artifacts | | | | | | | | social | | |
| --- | --- | --- | --- | --- | --- | --- | --- | --- | --- | --- | --- | --- | --- |
|  | work | under-standing | technical drawings | learning contents | inter-activity | soft-ware | hard-ware | machine tools | work-pieces | environ-ment | trai-ners | col-leagues | ob-servers |
| work | **97** | .072 | .216 | .052 | .289 | .072 | .072 | .247 | .134 | .031 | .072 | .010 | .010 |
| understanding | .259 | **27** | .222 | .148 | .370 | .185 | .000 | .111 | .037 | .000 | .148 | .037 | .111 |
| tech. drawings | .382 | .109 | **55** | .055 | .255 | .200 | .018 | .091 | .036 | .018 | .018 | .000 | .018 |
| learning contents | .152 | .121 | .091 | **33** | .333 | .182 | .000 | .000 | .030 | .030 | .030 | .000 | .000 |
| interactivity | .277 | .099 | .139 | .109 | **101** | .119 | .010 | .129 | .030 | .020 | .000 | .000 | .059 |
| software | .135 | .096 | .212 | .115 | .231 | **52** | .077 | .019 | .000 | .000 | .000 | .000 | .096 |
| hardware | .269 | .000 | .038 | .000 | .038 | .154 | **26** | .115 | .038 | .154 | .000 | .000 | .000 |
| machine tools | .649 | .081 | .135 | .000 | .351 | .027 | .081 | **37** | .027 | .000 | .054 | .000 | .027 |
| workpieces | .929 | .071 | .143 | .071 | .214 | .000 | .071 | .071 | **14** | .000 | .000 | .071 | .000 |
| environment | .333 | .000 | .111 | .111 | .222 | .000 | .444 | .000 | .000 | **9** | .000 | .000 | .000 |
| trainers | .778 | .444 | .111 | .111 | .000 | .000 | .000 | .222 | .000 | .000 | **9** | .000 | .000 |
| colleagues | .500 | .500 | .000 | .000 | .000 | .000 | .000 | .000 | .500 | .000 | .000 | **2** | .000 |
| observers | .100 | .300 | .100 | .000 | .600 | .500 | .000 | .100 | .000 | .000 | .000 | .000 | **10** |

For further analysis, the collected incidents have been classified. As presented in Table II, the main dimensions *process*, *artifacts* and *social* have been subdivided and refined. Entries on the diagonal, standing out in bold type, represent the total of incidents in the respective category. To enhance legibility, intersections of categories are presented as percentages. These are based on the total values of the main category on the left hand side. For instance, 97 incidents in total have been classified as part of work processes and 27 as part of understanding (or rather learning) processes. The intersection of both as number of incidents is 7, which results in $7/97 = .072$ in relation to all incidents of work processes and $7/27 = .259$ for understanding processes.

In addition to the observations, the apprentices provided 188 (after deduplication: 179) answers during the integrated survey. In the following, we present the results of the evaluation.

### A. Incidents related to work phases

In the following, we describe the negative and positive incidents with a strong relationship to activities in the individual work phases.

*Phase 1: Analysis of tasks:* In the first phase, two negative incidents occurred. In one case, displayed instructions have not been read and were only confirmed by pressing a button. In the other case, during working without LMS support, the needed technical drawing was missing and the required measure had to be extracted from a CAD model at a computer not directly at the machine tool. Nevertheless, the work with the CAD viewer was also rated strongly positive, as the individual displayed workpiece layers were presented in colors related to workpiece properties, such as required quality of metal surfaces to manufacture. The use of the 3D model for gradual analysis was also rated strongly positive in another case.

*Phase 2: Preparation of tools:* During the preparation, fitting tools have to be picked out, measured, and noted in a tooling sheet. In 3 cases, technical terms and abbreviations used in the graphical user interface for the sheet were not understood and therefore the incidents were rated negatively. As a result of this observation it was recommended to change the terms in the interface or augment them with a legend. In the preparation phase, apprentices in some companies have to change the location to cut blank piece of metal to approximate measures. In one company there was no Wi-Fi hotspot in the corresponding room which led to problems with data access. Possible solutions are caching functions like the HTML5 *Application Cache*, or the installation of additional hotspots. Compared to the work with paper sheets, dynamic lists in which positions of single fields can be changed and additional fields can be added during the planning process. All of this has been evaluated positively. Additionally, view changes between the zoomable technical drawing and the tooling sheet in the LMS worked without any problems, once the considered browser tab handling was understood.

*Phase 3: Documentation of upcoming process:* Apprentices define the production steps of the upcoming manufacturing process in the documentation phase. A negative incident was the forced repeated entry of tool names in the plan. This was handled by using the ditto mark, as it would be handled on paper. An enhancement of the used LMS is an auto-complete function for names of tools frequently used in single companies. Another negative incident was the use of the printed Mechanical and Metal Trades Handbook additional to the learning contents in the app. Although a combination of different learning materials would have been rated positively, the printed version of 2011 contained outdated values. Strongly positive incidents were problem-free and self-explanatory integration and operation of the LMS for the creation of workflow sheets.

Additionally, the use of the search function and the related immediate discovery of required information were rated as strongly positive for autonomous learning.

*Phase 4: Setup of machines:* Two negative incidents have been observed while apprentices were setting up machines in the traditional way, without using the app. In one case, the apprentice unconsciously lost the printed technical drawing and had to interrupt the working process. In the other case, the measurements of a blank piece of metal differed from the assumed measurements and the work plan had to be changed. Positive incidents concern the use of technical drawings. Without using the app, the drawing was printed and attached with a magnet to the machine. Also positively rated was the use of the digital drawing to set up the machine. In one case, a smartphone was used to take a photo of a screen, as no paper was available to print.

*Phase 5: Programming CNC machines:* When working without the app, an apprentice explained that he always writes the program at first and takes the measurements afterwards. While writing the program, he interrupts his work and takes measurements. That was an disordered sequence which took additional time. In contrast, a positive incident was the handling of a tablet computer, which was held in one hand while programming the machine without difficulty. In another case the measurements of the work plan have not been directly read, but copied from memory. As this is common practice and shows understanding, it was rated positively.

*Phase 6: Manufacturing:* During manufacturing, a machine crashed. As for an unknown reason it did not take a programmed piece of the program. In another case, the metal became useless due to a manufacturing error. That also appeared as the starting point was set incorrectly and in one more instance, the diameter was used instead of the radius, which also produced a useless workpiece. These negatively rated incidents occurred independently from the use of the LMS and have to be to be considered as usual in the learning process. A positive incident using the LMS was the use of the zoom function inside a PDF file containing a technical drawing. Additionally, the quality control of workpieces typically begins during the manufacturing, as single manufacturing steps are controlled after finishing. This is a main argument to show that the work phases are intertwined with each other. Consequently, the quality assurance sheet has to be available already at the manufacturing phase (see Fig. 1). This also highlights the importance of the interactive sheets as a link of required information among individual work phases.

*Phase 7: Quality control:* Strongly negative incidents during the quality control phase were two cases in which the LMS software prototype was not available during server errors (*HTTP 500 and 504*) as well as one case in which the control sheet was not added to the task and therefore not accessible. In contrast, in a positive incident during the control, a difference of the drawing and the produced workpiece came to attention, which would have had an important impact in operational production.

*B. Incidents in intersections of categories*

The classification of incidents using several categories per incident allows the examination of the intersections among the categories. We analyzed the individual categories in Table II, examined intersections with other categories, and extracted decisive events concerning processes, artifacts, and social interactions with regard to technical support of VET.

*1) Work processes and learning:* One of the most important resources for the work processes and learning are technical drawings, which are used in the entire working time. During the initial analysis, required measures are extracted and up to manufacturing and quality control, measures are compared with workpieces. This is regardless of the use of digital systems or conventional procedures. However, in the observations printed versions have been lost or the drawing became dirty and could not be read anymore. This was prevented by the use of tablets and notebooks, which have been handled with care. In addition, parts of digital drawings can be enlarged, which was rated as helpful.

Sheets are used to compile lists of tools needed for manufacturing individual workpieces, for planning the sequence of manufacturing steps, and for quality assurance of finished workpiece parts. Digital sheets of the LMS are displayed beginning from the respective work phase they are required until the end of a task. It was observed that integrated digital sheets, in which sequences of entries can be modified by drag and drop actions, provided relevant assistance in contrast to printed alternatives. Belated corrections and modifications of sheet entries could be carried out without problems. They were integrated sufficiently into overall task workflows and are highly related to the work with machine tools. However, some incidents indicate that simple lists of checkbox entries, which have been implemented to support the awareness of work steps that are sometimes omitted, have just been checked without reading them.

*2) Artifacts:* The category *Software* was used for the LMS itself as well as for observed user experience. A strongly positive rating was the use of an integrated 3D volume model of a workpiece for analysis. Also positive was the self-explanatory user interface (see Fig. 2), with which the trainees were quickly familiarized. Further incidents have already been described above.

The category *Hardware* was used to classify technical devices and installations for the LMS. Strongly negative ratings were a virtual keyboard overlapping a tablet screen as well as a room change with a following absence of WiFi connections. Additionally, in one incident work had to be interrupted due to an empty battery. The problem-free handling when typing on tablets was evaluated positively several times. Furthermore, additional Hardware like external keyboards and mouses were used without problems. Without efforts, tablets were sometimes held in one hand while typing. Loops on the protective cover of tablets have been helpful. Also docking stations for tablets, and notebooks, were used. The integrated camera of a tablet was utilized to take a photo of a screen when no paper was available for printing.

*3) Social interactions – questions and collaboration:* Trainers, apprentices, and developers in the project assumed that during the preparation phase, before the actual manufacturing, most questions from apprentices would come up. This was the result of a survey in a workshop of the project. Actually, the observations showed that 7 of 9 questions to trainers were asked during manufacturing with machines. Reasons for the questions were unknown handling of the machine controls, a non-fitting drill tool, mathematical conversions, a failed manufacturing step, (physical) handling of machines, and an incorrect configuration of a machine. The arise of these incidents during the manufacturing indicate a sufficient support of work processes during the preparation phases, as questions mainly arose related to the physical environment.

Although manufacturing tasks mainly consist of individual work, apprentices asked more experienced colleagues for help. This highlights the importance of the entire community for guidance at the workplace [5]. Additionally, the two observers were open to conversation throughout the entire time. Consequently, they were asked if apprentices had problems handling the LMS. Most of the questions concerned the best way to input data into the interactive protocols and general handling of the LMS (e.g. the use of browser tabs or keyboard shortcuts) in cases where the LMS was used for the first time. Although most of the apprentices agreed that the user interface was self-explanatory, a introductory workshop for any LMS is recommended.

## IV. DISCUSSION

The underlying project of this study and the developed mobile LMS aim to assist apprentices and trainers in established working methods. Special characteristics of VET in the dual system are the duration of the training of 3.5 years as well as learning and work tasks, which are applied during the entire learning time. Advantages of this study are direct cross-company observations of real-world tasks, the use of several observers with specialized domain knowledge, and the possibility of being involved in the first use of the LMS by apprentices. Limitations of the study are inevitably subjective during the rating of single observations and the use of a software prototype in beta state. Additionally, we were bound to time restrictions of everyday life in the training, with the result that not all observations could be accompanied by both of the two observers. Furthermore, the focus on the integration of the LMS led to a total number of 12 observations (233 incidents) with the LMS and 3 observations (37 incidents) without the LMS.

The most significant contribution of this project is the successful integration of the LMS into the task structure of the apprenticeship as illustrated in Fig. 1. There are two characteristics of the VET system considered that meet the needs of the support of a technical system. First, LWT can be subdivided into phases, which occur regularly. By splitting LWT into subtasks, the system can assist apprentices with information and appropriate technical features to achieve sub-targets. Second, tasks with similar structure are used throughout the entire period of the apprenticeship. In the metalworking domain, tasks become more complex over time. While tasks at the beginning of the training do not require programming, this phase can be added later when working with CNC machines. The LMS was successfully implemented in big companies with training facilities as well as in small training centers at inter-company level.

In summary, the following advantages of the mobile LMS resulted from the comparison with and without our system:

- Support in the adherence to work processes that build on each other.
- Understandable representation of 3D elements for beginners through interactive 3D models.
- Smooth enlargement of technical drawings through zoomable vector graphics.
- Interactive sheets with ability to perform subsequent corrections.
- The implementation of LMS user interfaces in the form of web applications enables the use of a wide range of technical devices, including input terminals of modern machine tools.
- Rapid distribution and consistent use of new versions of learning materials.
- Avoidance of the misplacement of handwritten notes and technical drawings.
- Autocompletion to avoid unnecessary time caused by repeated inputs (not implemented by now).

Equally, the following demands emerged and drawbacks resulted:

- Software errors can disrupt work processes.
- Workplaces should be equipped with WiFi or similar network connections.
- Properties of mobile devices have to be considered (battery life, required docking stations or external input devices).
- A training is required for the effective use of new software solutions.

For similar projects, we highly recommend a user-centered design. Results from the first interviews and conversations helped to avoid pitfalls in advance. For instance, we planned to implement the planning and documentation phase before the preparation phase, where machines and tools are assembled. In reality, however, it is not possible to predict which tools are actually available. Therefore, the preparation has to be placed before the documentation. A result of the observations of this study was the interconnection of the manufacturing phase and the quality control. Initially, the quality assurance sheet was available only in the quality control phase. However, the sheet is needed already during production, as measurements are taken during this time. The LMS has been adapted as a result.

In the future, we plan to integrate auto-completion features for the interactive tooling and workflow sheets. This feature has been asked for several times during the observations. A challenge is the handling of tracked periods of time. On the one hand, this is a feature for trainers to get an overview and perform comparisons of the work progression. On the other hand, it tends to restrict the rights of apprentices in terms of data protection and privacy.

The developed mobile learning system provides appropriate support in the field of metalworking, where special demands have to be considered. On the one hand, theoretical preparatory work is necessary to analyze characteristics of workpieces as well as to plan appropriate upcoming work steps and sequences. On the other hand, practical work is carried out on machines in different locations. In addition to these domain requirements, the implementation of the LMS was successful in several heterogeneous companies. These arguments support an extension of the described and applied underlying concepts in additional learning domains.